\documentstyle[psfig]{cupconf}

\ifoldfss
\else
  \ifnfssone
    \newmathalphabet{\mathit}
      \addtoversion{normal}{\mathit}{cmr}{m}{it}
      \addtoversion{bold}{\mathit}{cmr}{bx}{it}
    \newmathalphabet{\mathcal}
      \addtoversion{normal}{\mathcal}{cmsy}{m}{n}
    \else
    \ifnfsstwo
    \fi
  \fi
\fi

\def\kms{$\rm km\;s^{-1}$}
\def\msun{M$_{\odot}$}
\def\lsunV{$\rm L_{V_{\odot}}$}
\def\kmsmpc{$\rm km\;s^{-1}\;Mpc^{-1}$}
\def\ml{(M/L)$_{\odot}$}
\def\ha{H$\alpha$}
\def\nii{[N~{\scriptsize II}]}
\def\oii{[O~{\scriptsize II}]}

\def\hexnumber#1{\ifcase#1 0\or1\or2\or3\or4\or5\or6\or7\or8\or9\or
 A\or B\or C\or D\or E\or F\fi }

\makeatletter
\ifx\CUP@mtlplain@loaded\undefined
\else
\fi
\makeatother
 \makeatletter
 \ifx\CUP@mtlplain@loaded\undefined
   \font\tenbmi=cmmib10 at 10pt
   \font\sevenbmi=cmmib10 at 7pt
   \font\fivebmi=cmmib10 at 5pt

   \newfam\bmifam
   \textfont\bmifam=\tenbmi
   \scriptfont\bmifam=\sevenbmi
   \scriptscriptfont\bmifam=\fivebmi
   
 \fi
 \makeatother
\ifnfsstwo

\fi
\ifnfssone

\fi
\ifoldfss

\fi

\mathchardef\varLambda="0103

\makeatletter
\ifx\CUP@mtlplain@loaded\undefined
\else
\fi
\makeatother
\makeatletter
\ifx\CUP@mtlplain@loaded\undefined
  \font\tenbms=cmbsy10
  \font\sevenbms=cmbsy10 at 7pt
  \font\fivebms=cmbsy10 at 5pt
  \newfam\bmsfam
  \textfont\bmsfam=\tenbms
  \scriptfont\bmsfam=\sevenbms
  \scriptscriptfont\bmsfam=\fivebms

  \edef\bsy@{\hexnumber\bmsfam}
  \mathchardef\bnabla="0\bsy@72
\fi
\makeatother
\title[The kinematics and the origin of the ionized gas in NGC~4036]{
The kinematics and the origin of the ionized gas in NGC~4036}

\author[E.M. Corsini {\it et al.\/}]%
{E.\ns M.\ns C\ls O\ls R\ls S\ls I\ls N\ls I$^1$,\ns
 F.\ns B\ls E\ls R\ls T\ls O\ls L\ls A$^1$,\ns
 M.\ns S\ls A\ls R\ls Z\ls I$^1$,\ns\\
 P.\ns C\ls I\ls N\ls Z\ls A\ls N\ls O$^1$,\ns
 H.\ls-\ls W.\ns R\ls I\ls X$^2$\ns 
 \and \ns 
 W.\ls W.\ns Z\ls E\ls I\ls L\ls I\ls N\ls G\ls E\ls R$^3$}

\affiliation{$^1$Dipartimento di Astronomia, Universit\`a di Padova,
Vicolo dell'Osservatorio 5, I-35122 Padova, Italy\\[\affilskip]
$^2$Steward Observatory, University of Arizona, Tucson, AZ-85721\\[\affilskip]
$^3$Institut f\"ur Astronomie, Universit\"at Wien, T\"urkenschanzstrasse 
17, A-1180 Wien, Austria}

\begin{document}
\ifnfssone
\else
  \ifnfsstwo
  \else
    \ifoldfss
      \let\mathcal\cal
      \let\mathrm\rm
      \let\mathsf\sf
    \fi
  \fi
\fi

\maketitle

\begin{abstract}
The kinematics of stars and ionized gas has been studied near the
center of the S0 galaxy NGC~4036.  Dynamical models based both on
stellar photometry and kinematics have been built in order to derive
the gravitational potential in which the gas is expected to orbit.
The observed gas rotation curve falls short of the circular velocity
curve inferred from these models. Inside $10''$ the observed gas
velocity dispersion is found to be comparable to the predicted
circular velocity, showing that the gas cannot be considered on
circular orbits.  The understanding of the observed gas kinematics is
improved by models based on the Jeans Equations, which assume the
ionized gas as an ensemble of collisionless cloudlets distributed in a
spheroidal and in a disk component.
\end{abstract}

\firstsection 

\section{Introduction}

NGC~4036 has been classified S0$_{3}$(8)/Sa in RSA (Sandage
\& Tammann 1981) and S0$^{-}$ in RC3 (de Vaucouleurs {\it et al.\/} 
1991). Its total apparent magnitude is $V_T\,=\,10.66$ mag (RC3).
This corresponds to a total luminosity $L_V\,=\,4.2 \cdot 10^{10}$
\lsunV\ at the assumed distance of $d\,=\,V_0/H_0\,=\,30.2$ Mpc, where
$V_0\,=\,1509\pm50$ \kms\ (RSA) and assuming $H_0\,=\,50$ \kmsmpc.  At
this distance the scale is 146 pc arcsec$^{-1}$.

We measured the kinematics of stars and ionized gas along the galaxy
major axis and we also derived their distribution in the nuclear
regions by means of ground-based $V-$band and HST narrow-band
imaging respectively.

\section{Modeling the stellar kinematics}

We apply to the observed stellar kinematics the Jeans modeling
technique introduced by Binney {\it et al.\/} (1990), developed by van der
Marel {\it et al.\/} (1990) and van der Marel (1991), and extended to
two-component galaxies by Cinzano \& van der Marel (1994).

The best-fit model to the observed major-axis stellar kinematics is
shown in Fig.~1. The bulge is an oblate isotropic rotator ($k=1$) with
$(M/L_V)_b$ = 3.4 \ml. The exponential disk has $(M/L_V)_d$ = 3.4 \ml\
with the velocity dispersion profile given by
$\sigma_\star(r)\,=\,155\;e^{-r/r_\sigma}$ \kms\ with scale-length
$r_{\sigma}\,=\,27.4''\,=\,4.0$ kpc.  The derived bulge and disk
masses are $M_{b}\,=\,9.8 \cdot 10^{10}$ \msun\ and $M_{d}\,=\,4.8
\cdot 10^{10}$ \msun , adding up to a total (bulge$+$disk) luminous
mass of $M_T\,=\,14.5 \cdot 10^{10}$ \msun.  The disk-to-bulge and
disk-to-total luminosity ratios are $L_b/L_d\,=\,0.58$ and
$L_d/L_T\,=\,0.36$.

\begin{figure}
\vspace{8.2cm}
\includegraphics{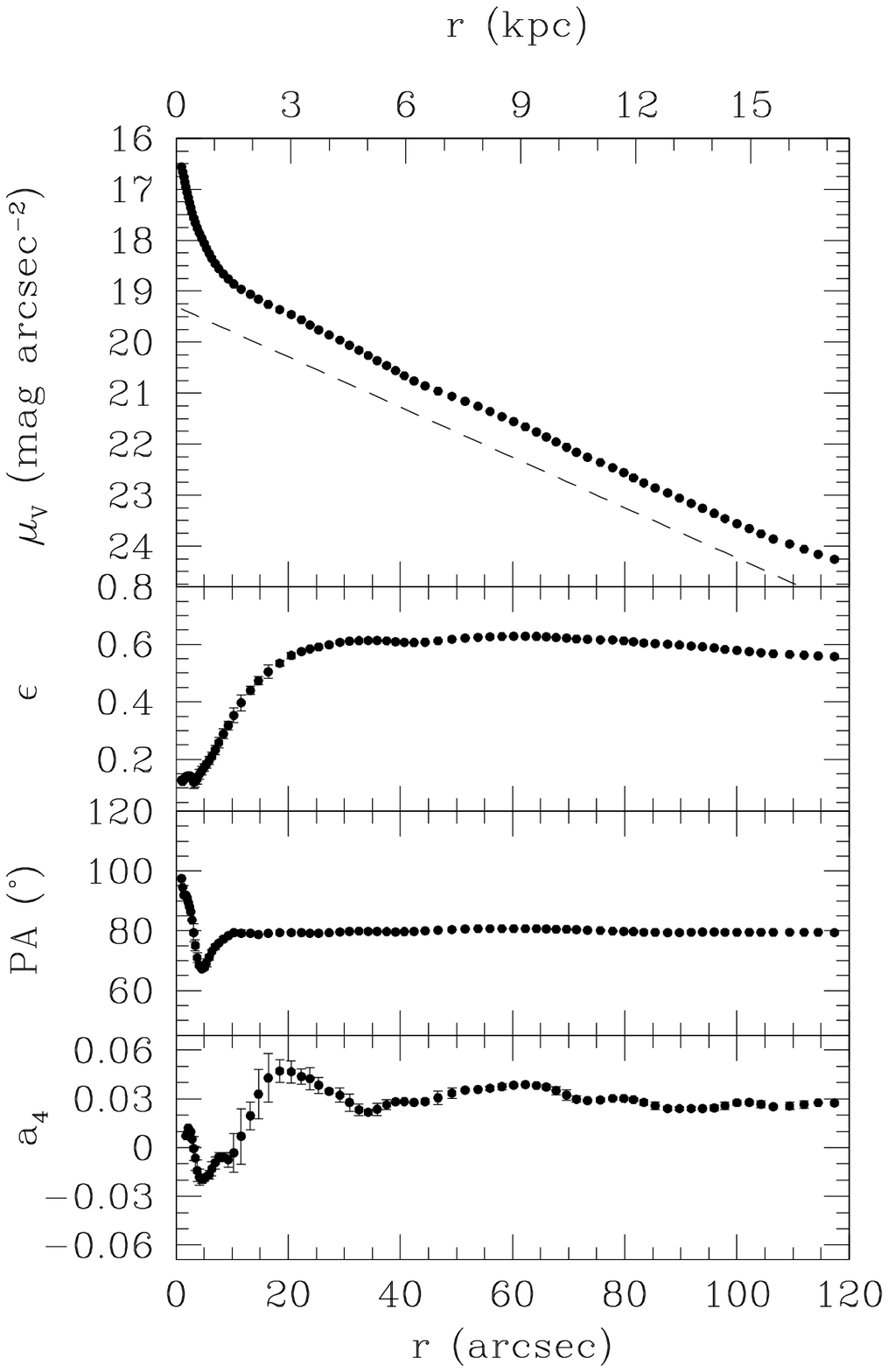}
\includegraphics{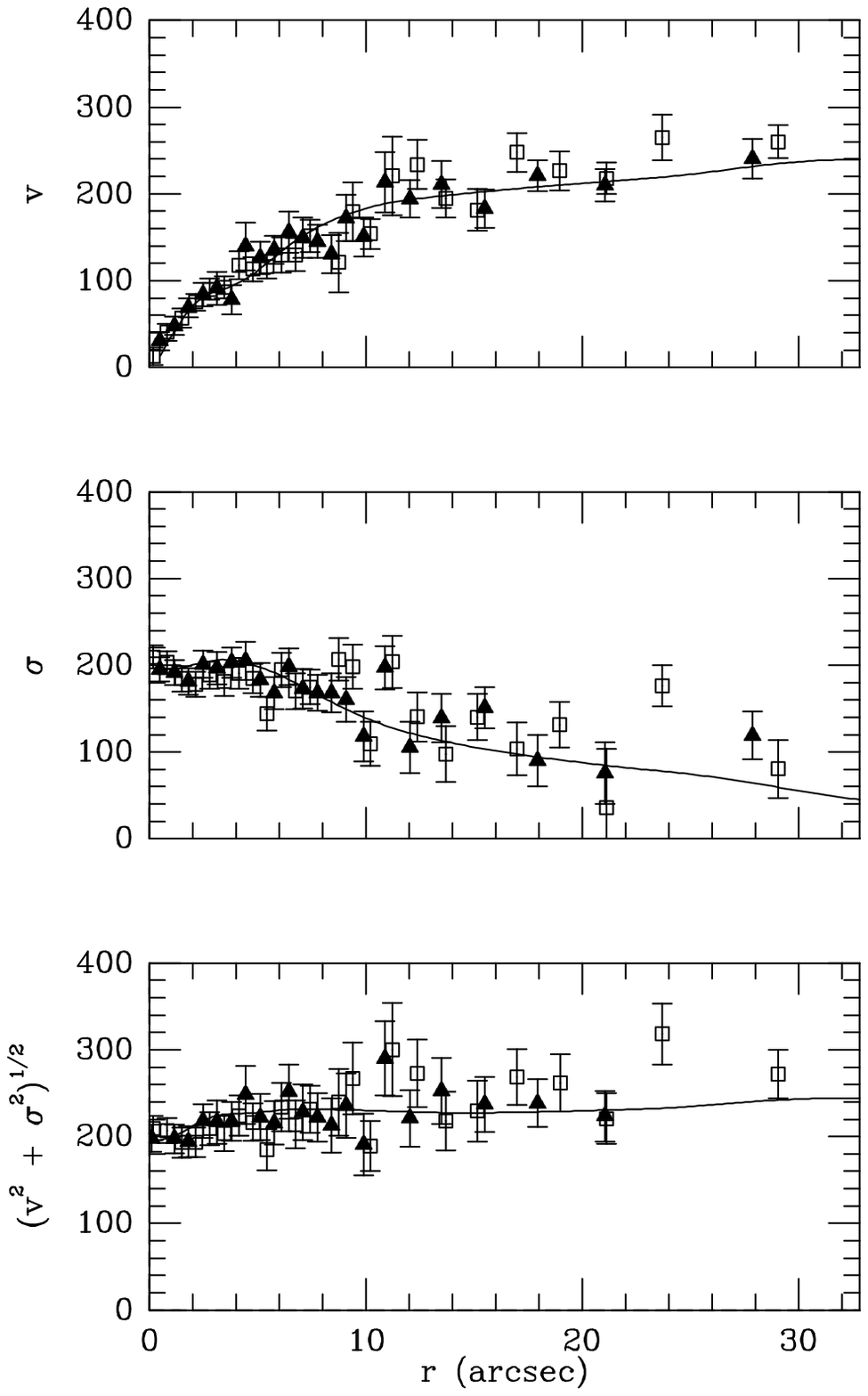}
\caption{{\it Left panel:\/} Ground-based $V-$band surface-brightness,
ellipticity, position angle and $\cos 4\theta$ coefficient radial
profiles of NGC~4036. The dashed line represents the
surface-brightness profile of the exponential disk adopted in our
dynamical model.  {\it Right panel:\/} Comparison of the model
predictions to the stellar major-axis kinematics for NGC 4036}
\end{figure}

\section{Modeling the gaseous kinematics}

At small radii both the ionized gas velocity and velocity dispersion
are comparable to stellar velocity and velocity dispersion, for
$r\leq9''$ and $r\leq5''$ respectively. Moreover a change in the slope
of the \oii$\,\lambda3726$ intensity radial profile is observed
inside $r\simeq8''$, its gradient appears to be somewhat steeper
towards the center. The velocity dispersion and intensity profiles of
the ionized gas suggest that it is distributed into two components: a
small inner spheroidal component and a disk.  We decomposed the
\oii$\,\lambda3726$ intensity profile as the sum of an $R^{1/4}$
gaseous spheroid and an exponential gaseous disk and the gas spheroid
resulted to be the dominating component up to $r\simeq8''$.

We built up dynamical models for the ionized gas in NGC~4036
(Fig.~2). It was assumed to be distributed in a dynamically hot
spheroidal and in a dynamically cold disk component and consisting of
collisionless individual clumps (cloudlets) which orbit in the total
potential. We made two different sets of assumptions based on two
different physical scenarios for the gas cloudlets.\\
{\it Model A\/}: In a first set of models we described the gaseous
component consisting of collisionless cloudlets which can be
considered in hydrostatic equilibrium. The gaseous spheroid is
characterized by a density distribution and flattening different from
those of stars. Its major-axis luminosity profile was assumed to
follow an $R^{1/4}$ law.  The flattening of the spheroid $q$ was kept
as free parameter.  To derive the kinematics of the gaseous spheroid
and disk we solved the Jeans Equations.\\
{\it Model B\/}: In a second set of model we assumed that the emission
observed in the gaseous spheroid and disk arise from material that was
recently shed from stars. Different authors (Bertola {\it et al.\/}
1984, 1995b; Fillmore {\it et al.\/} 1986; Kormendy \& Westpfahl 1989;
Mathews 1990) suggested that the gas lost (e.g. in planetary nebulae)
by stars was heated by shocks to the virial temperature of the galaxy
within $10^4$ years, a time shorter than the typical dynamical time of
the galaxy. Hence in this picture the ionized gas and the stars have
the same true kinematics, while their observed kinematics are
different due to the line-of-sight integration of their different
spatial distribution.

\begin{figure}
\vspace{5.5cm}
\includegraphics{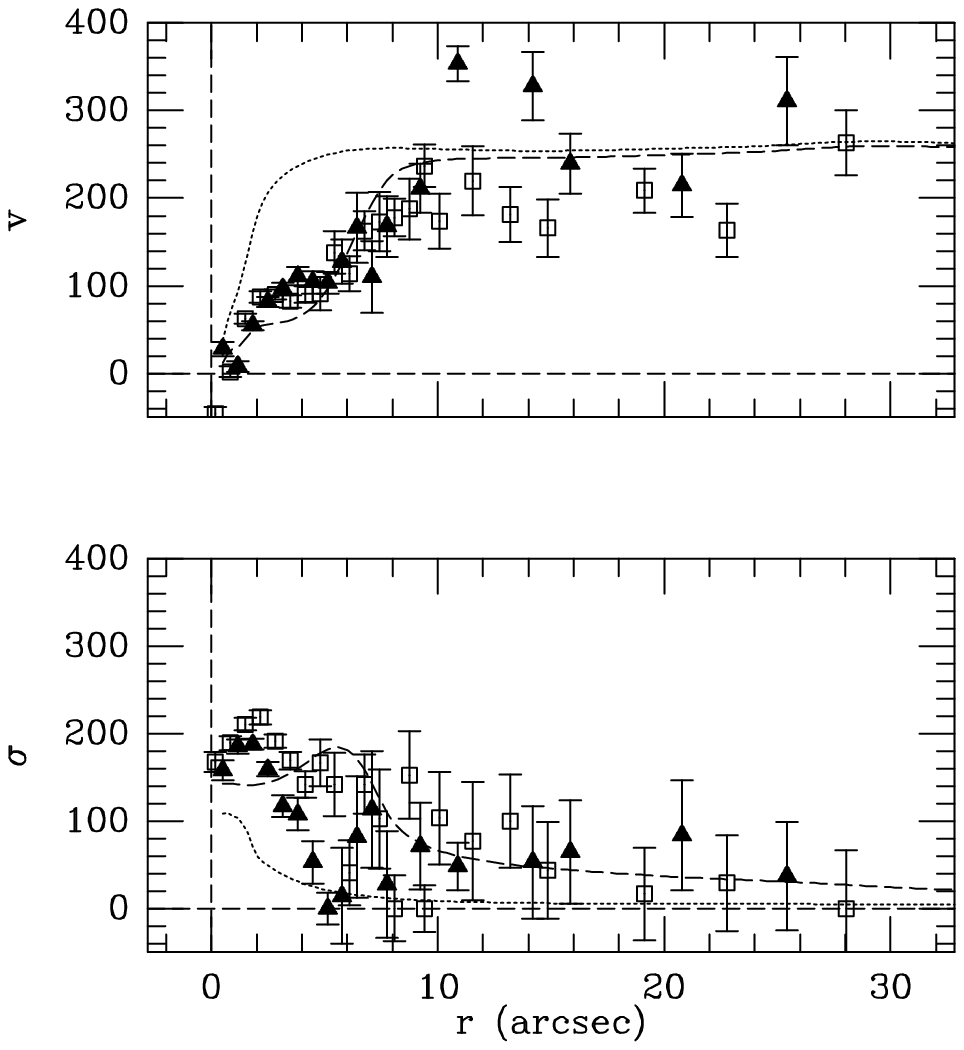}
\includegraphics{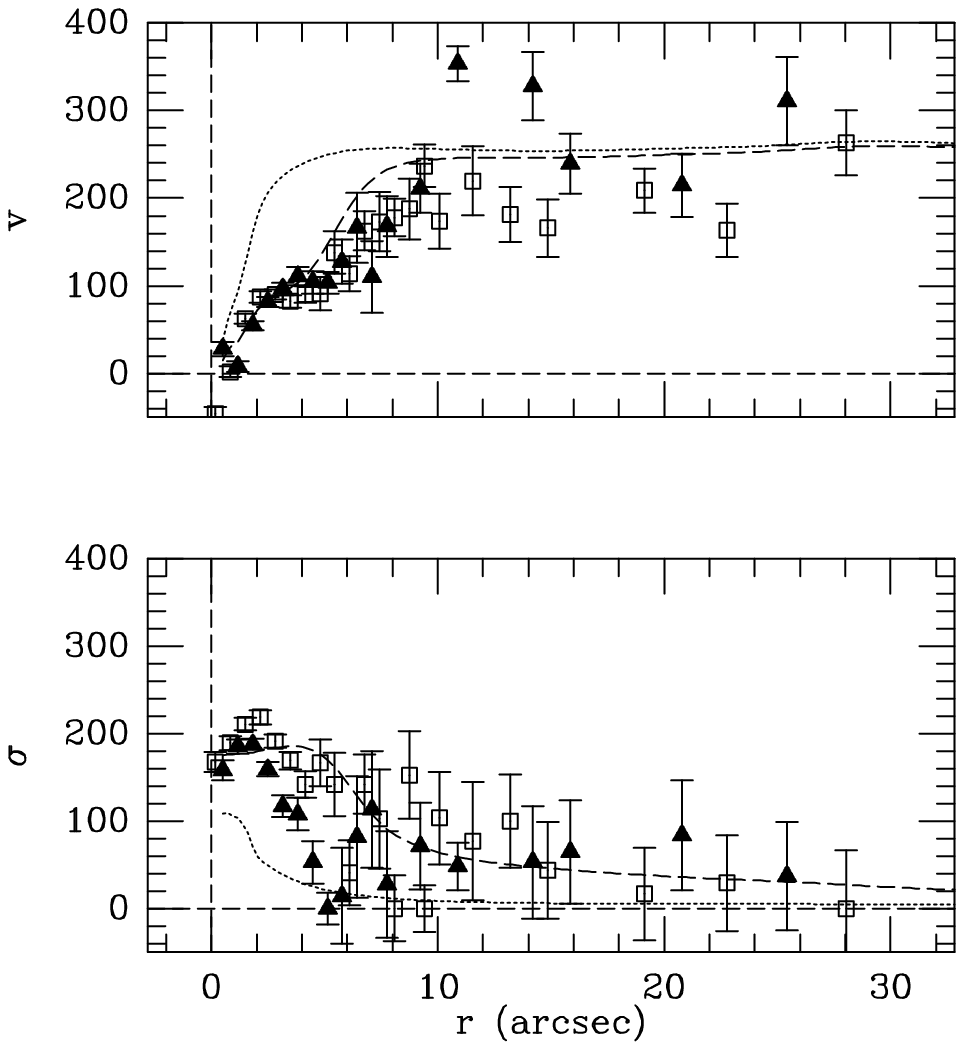}
\caption{Comparison of the predictions (dashed curves) of model A
({\it left panel\/}) and model B ({\it right panel\/}) to the ionized
gas kinematics observed along the major axis of NGC~4036.  The dotted
curves represent the seeing-convolved circular velocity curve and zero
velocity-dispersion profile in the galaxy meridional plane
respectively}
\end{figure}

\section{Do drag forces affect the kinematics of the gaseous cloudlets?}

The discrepancy between model and observations could be explained by
taking properly into account the drag interaction between the ionized
gas cloudlets of the gaseous spheroid and the hot component of the
interstellar medium (Mathews 1990). To have some qualitative insights
in understanding the effects of a drag force on the gas kinematics we
studied the case of a gaseous nebula moving in the spherical potential
generated by an homogeneous mass distribution of density $\rho$ and
which, starting onto a circular orbit, is decelerated by a drag force
${\bf F_{\it drag}}\,=\,-(k_{\it drag}v/m)\,{\bf v}$, where $m$ and
${\bf v}$ are the mass and the velocity of the gaseous cloud and the
constant $k_{\it drag}$ is given following Mathews (1990).  We
numerically solved the equations of motion of a nebula to study the
time-dependence of the radial and tangential velocity components
$\dot{r}$ and $r\dot{\psi}$. We fixed the potential assuming a
circular velocity of 250 \kms\ at $r = 1$ kpc.  Following Mathews
(1990) we took an equilibrium radius for the gaseous nebula $a_{\it
eq}=0.37$ pc.  It results that $\ddot{\psi}<0$ and $\ddot{r}>0$: the
clouds spiralize towards the galaxy center as expected. Moreover the
drag effects are greater on faster starting clouds and therefore
negligible for the slowly moving clouds in the very inner region of
NGC~4036.

If the nebulae are homogeneously distributed in the gaseous spheroid
only the tangential component $r\dot{\psi}$ of their velocity
contribute to the observed velocity.  No contribution derives from the
radial component $\dot{r}$ of their velocities.  In fact for each
nebula moving towards the galaxy center, which is also approaching to
us, we expect to find along the line-of-sight a receding nebula, which
is falling towards the center from the same galactocentric distance
with an opposite line-of-sight component of its $\dot{r}$.  However
the radial components of the cloudlets velocities (typically of 30-40
\kms) are crucial to explain the velocity dispersion profile and to
understand how the difference between the observed velocity
dispersions and the model B predictions arises. If the clouds are
decelerated by the drag force their orbits become more radially
extended and the velocity ellipsoids acquire a radial anisotropy.

So we expect that (in the region of the gaseous spheroid) including
drag effects in our gas modeling should give a velocity dispersion
profile steeper than the one predicted by our isotropic model B, and
in better agreement with observations.

\section{Discussion and conclusions}

The modeling of the stellar and gas kinematics in NGC~4036 shows that
the observed velocities of the ionized gas, moving in the
gravitational potential determined from the stellar kinematics, can
not be explained without taking the gas velocity dispersion into
account. In the inner regions of NGC~4036 the gas is not moving at
circular velocity.

A better match with the observed gas kinematics is found by assuming
the ionized gas as made of collisionless clouds in a spheroidal and
disk component for which the Jeans Equations can be solved in the
gravitational potential of the stars (i.e., model A). A much better
agreement is achieved by assuming that the ionized gas emission comes
from material which has recently been shed from the bulge stars (i.e.,
model B). If this gas is heated to the virial temperature of the
galaxy (ceasing to produce emission lines) within a time much shorter
than the orbital time, it shares the same `true' kinematics of its
parent stars. If this is the case we would observe a different
kinematics for ionized gas and stars due only to their different
spatial distribution.  An HST \ha$+$\nii\ image of the nucleus of
NGC~4036 confirms that except for a complex emission structure inside
$3''$ the smoothness of the distribution of the emission as we expect
for the gas spheroidal component.

This kinematical modeling leaves open the questions about the physical
state (e.g. the lifetime of the emitting clouds) and the origin of the
dynamically hot gas. We tested the hypothesis that the ionized gas is
located in short-living clouds shed by evolved stars (e.g. Mathews
1990) finding a satisfying agreement with our observational
data. These clouds may be ionized by the parent stars, by shocks, or
by the UV-flux from hot stars (Bertola {\it et al.\/} 1995a).  The
comparison with the more recent and detailed data on gas by Fisher
(1997) opens wide the possibility for further modeling improvement if
the drag effects on gaseous cloudlets (due to the diffuse interstellar
medium) will be taken into account.  These arguments indicate that the
dynamically hot gas in NGC~4036 has an internal origin. This does not
exclude the possibility for the gaseous disk to be of external origin
as discussed for S0's by Bertola {\it et al.\/} (1992).  Spectra at
higher spatial resolution are needed to understand the structure of
the gas inside $3''$.


\begin{thebibliography}{}

\bibitem[Bertola {\it et al.\/} 1984]{ber84}{\sc Bertola,
    F., Bettoni, D., Rusconi, L. \& Sedmak, G.} 1984, {\it AJ} {\bf 89}, 356

\bibitem[Bertola {\it et al.\/} 1992]{ber92}{\sc Bertola, F.,
    Buson, L.M. \& Zeilinger, W.W.} 1992, {\it ApJ} {\bf 401}, L79

\bibitem[Bertola {\it et al.\/} 1995a]{ber95a}{\sc Bertola, F., Bressan, A., 
    Burstein, D. et al.}
    1995, {\it ApJ} {\bf 438}, 680

\bibitem[Bertola {\it et al.\/} 1995b]{ber95b}{\sc Bertola, F., Cinzano, 
    P., Corsini, E.M., Rix, H.-W. \& Zeilinger, W.W.} 
    1995, {\it ApJ} {\bf 448}, L13

\bibitem[Binney {\it et al.\/} 1990]{bin90}{\sc Binney, J.J.,
    Davies, R.L. \& Illingworth, G.D.} 1990, {\it ApJ} {\bf 361},78

\bibitem[Cinzano \& van der Marel 1994]{cin94}{\sc Cinzano, P. \& van der
    Marel, R.P.} 1994, {\it MNRAS} {\bf 270}, 325

\bibitem[de Vaucouleurs {\it et al.\/} 1991]{RC3}{\sc de Vaucouleurs, G.
    et al.}
    1991, Third Reference Catalogue of Bright Galaxies.
    Springer-Verlag, New York (RC3)

\bibitem[Fillmore {\it et al.\/} 1986]{fil86}{\sc Fillmore, J.A.,
    Boroson, T.A. \& Dressler, A.} 1986, {\it ApJ} {\bf 302}, 208

\bibitem[Fisher 1997]{fis97}{\sc Fisher, D.} 1997, {\it AJ} {\bf 113}, 950

\bibitem[Kormendy \& Westpfahl 1989]{kor89}{\sc Kormendy, J. \& Westpfahl,
    D.J.} 1989, {\it ApJ} {\bf 338}, 772

\bibitem[Mathews, W.G.]{mat90}{\sc Mathews, W.G.} 1990, {\it ApJ} {\bf 354}, 468

\bibitem[Sandage \& Tammann 1981]{RSA}{\sc Sandage, A. \& Tammann, G.A.}
    1981, A Revised Shapley--Ames Catalog of Bright Galaxies. 
    Carnegie Institution, Washington (RSA)

\bibitem[van der Marel, R.P.]{vdm91}{\sc van der Marel, R.P.} 1991, {\it
    MNRAS} {\bf 253}, 710

\bibitem[van der Marel {\it et al.\/} 1990]{vdm90}{\sc van der
    Marel, R.P., Binney, J.J. \& Davies, R.L.} 1990, {\it MNRAS} {\bf 245}, 582


\end{thebibliography}
\end{document}